# Modulated Collective Motions and Condensation of Bacteria

Mei-Mei Bao(鲍美美)[1†], Isaiah Eze Igwe[2†], Kang Chen(陈康)[1*], and Tian-Hui Zhang(张天辉)[1*]

[1]*Center for Soft Condensed Matter Physics and Interdisciplinary Research & School of Physical Science and Technology, Soochow University, Suzhou 215006, China*

[2] *Department of Physics, Federal University Dutsin-Ma, Katsina State 821101, Nigeria*

*†These authors contributed equally to this work.*

*\*Correspondence to zhangtianhui@suda.edu.cn*

PACS: 87.80.-y;47.57.jd;87.15.Zg;87.17.-d;

**Abstract**

Bacteria can spontaneously develop collective motions by aligning their motions in dense systems. Here, we show that bacteria can also respond collectively to an alternating electrical field and form dynamic clusters oscillating at the same frequency of the field. As the dynamic clusters go beyond a critical size, they split into smaller ones spontaneously. The critical size for splitting depends on the frequency of electric field and the concentration of bacteria. We show that instead of their biological activity, the physical properties of bacteria as charged particles are responsible for the formation of dynamic clusters. Electroconvective flows across the system play the key role in stabilizing the clusters. However, to form clusters, collective hydrodynamic cooperation between bacteria is important such that no aggregation occurs in dilute suspensions. The findings in this study illustrate that bio-systems can respond collectively to an external field, promising an effective way to control and modulate the behavior of organisms. Moreover, the controlled aggregation and condensation of bacteria offer a robust approach to improve the local concentration of bacteria for early and rapid detection, which has wide applications in clinics.

**Introduction**

Bacteria can form various collective motions [1-3]. In these observations, hydrodynamic interactions between bacteria play a key role in aligning velocity for collective motions. Collective motions including flocking and rotating vortices have been observed and studied extensively [4-7]. In this case, collective motions can be successively directed and modulated by applying control on hydrodynamic flows[8-15]. For example, as dense suspensions of *Bacillus subtilis* were confined in a long and narrow macroscopic 'racetrack' channel, spiral vortices emerged [16]. In circular channels, the geometrical confinement on flows gave rise to directed fluid flows, and then arrays of counter-rotating vortices formed [16].

In suspensions of magnetotactic bacteria (MTB), the collective behaviors can be tuned and modified by an external magnetic field due to the interplay between the hydrodynamic interactions and the field-induced interactions [11, 17, 18]. As MTB were confined in droplets, vortex and patterns were achieved [19, 20]. These results demonstrate that the combination of hydrodynamic interactions and the field-induced interactions can produce much more complex collective behaviors. However, magnetic field is applicable only for magnetotactic bacteria. As a popular property, bacteria, such as *E. coli*, are generally negatively charged in suspension [21, 22]. It is expected that the coupling between hydrodynamics and an electric field may impose additional control on the behavior of individual bacteria and thus their collective behaviors. As an effective method, electric fields have been employed in separating and trapping cells. For example, blood cells characterized by different electrophoretic properties can be separated and selected by an electric field [23, 24]. If the electric field is spatially patterned, cells can be trapped and formed patterned distribution [25]. Despite of its wide interest, however, the effect of electric field on collective motions of bacterial has not been explored.

Here, in this study, the collective behaviors of *E. coli* bacteria were investigated under an alternating current (AC) field. It is found that the *E. coli* bacteria can respond collectively to a low-frequency field and form dynamic clusters. The radius of dynamic clusters oscillates with the field at the same frequency. They diffuse and merge into larger ones as they meet. However, the size of the cluster is self-limited: as the radius of the clusters exceeds a critical size, they split into smaller ones. The field-induced reversible formation of oscillating dynamic clusters demonstrates a new type of collective behaviors of active matter. Moreover, the formation of dynamic clusters of bacteria can significantly improve the local concentration of bacteria. Bacterial infections are currently a major cause of fatal disease in humans. Rapid and accurate detection of bacteria is the key to prevent large-scale infectious diseases [26]. Traditional diagnosis and characterization techniques need to incubate bacterial specimens from hours to days to increase concentration up to detectable levels. The reversible aggregation and condensation of active bacteria observed in this study provide a robust approach to quickly improve the local concentration of bacteria.

**Experimental Methods**

Bacteria (*E. coli*, CGMCC 1.12883) is obtained from the China General Microbiological Culture Collection. The selected bacterial colony is cultivated in a beef extract peptone medium [beef extract 0.3% (w/v), tryptone 1% (w/v), and sodium chloride 0.5% (w/v)] for 18 hours at 37°C in an incubator shaken at 150 rpm. The bacterial culture is centrifuged for 10 minutes (Minispin, Eppendorf) at 3000 rpm to remove the culture solution. The bacterial concentration is then diluted to an optical density around 1.0 which is measured at the wavelength of 600 nm by a microplate reader (Infinite F50, TECAN). The bacterial culture is rinsed with distilled water for three times. The final bacterial suspensions are prepared with a concentration of $2.4 \times 10^8$ cfu/ml (colony formation unit, indicating the total number of bacterial communities per milliliter) [27]. To acquire a full three-dimensional (3D) view of the dynamic clusters, the bacteria are dyed with CFDA-SE (Carboxyfluorescein diacetate succinimidyl ester) (400 nM, Sigma). The 3D dynamic clusters in the final steady state were scanned by confocal laser scanning microscopy (CLSM). To examine the role of bacteria mobility, similar observations are conducted with passive colloidal particles (Thermo Scientific G1000, polystyrene spheres) with a diameter of 2.0 $\mu$m.

**Results**

The bacterial suspension was sealed between two conducting indium-tin-oxide (ITO) coated glasses which are separated by insulating separators (Fig. 1a). The height of the insulating separator is around 180 $\mu$m. The dynamic processes are observed by a Nikon microscope with the 40x objective and recorded by a CMOS camera (CB019MG-LX-X8G3) at a rate of 10 frames per second. *E. coli* cells are generally negatively charged [28]. The cellular surface charge of *E. coli* cells has been quantified by a zeta potential of -10 ~ -40 mV [28]. However, the exact value of zeta potential varies between species and strains [28, 29]. To avoid directed motions along the field *E*, an AC field is employed (Fig. 1a). In this study, all observations were conducted with a bacterial concentration of $2.4 \times 10^8$ cfu/ml if it is not mentioned specially. This concentration is far below the critical concentration for flocking and swarming [2, 30] such that no collective motions can occur spontaneously in the absence of an electric field. Before the application of AC field, bacterial swim randomly and distribute uniformly. When the peak strength $E_p$ of AC is above 0.05 V/$\mu$m, bacterial cells are absorbed quickly to the electrodes and got damaged there. Therefore, $E_p$ in this study is maintained below 0.05 V/$\mu$m. As the frequency of AC field is above 100 Hz, no response and aggregation of bacteria are observed. At frequencies between 1.0~100 Hz, the system separates into two coexisting phases: a dense phase and a spare phase (Movie 1). The dense phase is not continuous but consists of interconnected dense domains. The dense domains are in dynamic equilibrium with the spare phase: Individual bacteria exchange between these two phases. When the suspension is subjected to a field with a frequency below 1.0 Hz, the dense domains collapse and form compact oscillating clusters.

Figures 1b-d present a dynamic process observed at the field of 0.035 V/$\mu$m and 0.2 Hz (Movie 2). As the field is switched on, dense domains emerge after an induction time (~90 s) (Fig. 1b). Three-dimensional (3D) clusters form subsequently within the dense domains (Fig. 1c). The resulting clusters are dynamic: they diffuse and merge into larger ones as they meet (Movie 3). However, the growth of clusters is self-limited: as the size of clusters is beyond a critical value, they split into smaller ones (Movie 4). Therefore, the clusters don't coalesce into a homogenous bulk phase but separate from each other with a finite size, giving rise to a discrete dense phase (Fig. 1d).

Three-dimensional Confocal scanning (see Methods) shows that the dynamic clusters are spherical (Fig. 1e). They suspend in the solvent and are not in contact with electrodes. Around the dense clusters, a sparse phase of diffusing bacteria can be identified. The sparse phase and the clusters are in dynamic equilibrium as can be seen in Movie 3. It follows that the formation of dense dynamic clusters is a result of microphase separation instead of a macroscopic phase separation. Inside the clusters, bacteria are active and move randomly (Movie 5). No spatial ordering is identified. As we switched off the electric field, bacteria begin to swim away from the clusters. As a result, the dynamic clusters dissociate gradually, and bacteria become free. It follows that the bacteria in the clusters are biologically active.

Since the bacteria are negatively charged, to maintain the dynamic clusters, the electrostatic repulsion between bacteria has to be balanced by an additional force. In colloidal suspensions subjected to an electric field, it has been well-known that a long-range attraction induced by the electrohydrodynamic (EHD) flows can bring charged colloidal particles together to form two-dimensional colloidal crystals [31]. However, the EHD-induced long-range attractions work mainly near electrodes. It is unlike that the dynamic clusters in the bulk are maintained by the EHD-induced long-range attraction. Except the EHD flows, electroconvective flows across the liquid offer a possible mechanism for the formation of dynamic clusters as sketched in Fig. 1f. Both theoretical and experimental studies have shown that as a dielectric liquid is subjected to an electric field, electroconvective flows across the bulk of liquid will form [32, 33]. However, as EHD flows near the surface of electrodes can be identified and visualized by tracer particles [34], it failed to demonstrate the three-dimensional electroconvective flows by tracer particles which are moving and disappear quickly from the focus plane [33]. Alternatively, fluorescent dye molecules were dispersed in the dielectric liquid. It was found that as the electroconvective flows emerge in the system, the intensity of fluoresce becomes patterned in space [33]. Nevertheless, if the gravity of colloidal particles is negligible, charged colloidal particles dispersed in the dielectric liquid will be transported to electrodes by electrophoretic forces [35]. Therefore, electroconvective flows in the bulk don't interact with colloidal particles, and no aggregation occurs. By contrast, as the gravity becomes significant, colloidal particles can be levitated and maintained in the bulk of liquid because of the balance between electrophoretic forces and the gravity [35]. The levitated particles can be organized into bicontinuous patterns by the electroconvection rolls [36]. The bicontinuous patterns of colloidal particles served a strong evidence of electroconvection flows [34, 37-40].

Here, we suggest that the dynamic clusters of bacteria are maintained by the electroconvective rolls as illustrated in Fig. 1f. In this case, the bacteria behave like charged colloidal particles, and their biological activity is not important in the formation of clusters. The difference is that bacteria are active and the effect of gravity can be neglected. Therefore, three-dimensional (3D) structures can form directly in the bulk of liquid without a biased electric field to balance the gravity. To verify the understanding on the role of biological activity, observations were conducted with the suspension of polystyrene (PS) particles of 2.0 $\mu$m diameter. PS particles (1.05 g/cm$^3$) have the similar density of water such that the gravity can be balanced by the buoyant force. As PS particles are dispersed in water, they become charged [31], and should exhibit similar clustering phenomena as bacteria if the biological activity has no effect on the formation of dynamic clusters. As expected, as the colloidal suspension is subjected to a field of 0.035 V/$\mu$m and 0.3 Hz, dynamic clusters form as well (Movie 6). Both merging and splitting of clusters are observed. It follows that the formation of dynamic clusters is mainly induced by the physical properties of charged particles and the biological activity is not critical.

To examine the role of individual dynamics in the formation of clusters, individual behaviors under the AC field are observed with bacteria and PS colloidal particles respectively in dilute suspensions. Both bacteria (Movie 7) and PS colloidal particles (Movie 8) move up and down periodically with the oscillating electric field. No aggregation and clustering take place. This is in consistent with previous studies at constant electric fields [35, 36]. It follows that collective cooperation between charged particles and their electrohydrodynamic interactions are essential in forming the macroscopic clusters. Different from passive colloidal particles, bacteria are active. As a consequence, it is found that as individual bacteria move up and down with the AC field, they also exhibit diffusive displacements in-plane (perpendicular to the electric field) (Movie 7). However, the displacements are at the same scale of bacteria size, being far smaller than the dimension of convection rolls. Therefore, as active particles, bacteria may disturb local flows slightly at the scale of microscopic meter. At larger scale, their collective behavior as charged particles becomes dominant.

At constant fields, electrophoretic forces transport charged particles to oppositely charged electrode. In this case, the gravity of particles is important to balance the electrophoretic forces and maintain particles in the bulk of liquid. For bacteria or buoyant particles, periodically inverted electric fields provide an alternative approach to keep them in the bulk of suspension: As the magnitude and the frequency of AC field are well matched, particles moving toward oppositely charged electrodes will be pulled back by the periodically inverted electrophoretic forces before they reach the electrodes as can be seen clearly in Movie 7 and Movie 8. Moreover, as the direction of AC field inverts periodically, the magnitude of electric field changes continuously and periodically. As a consequence, the magnitude of the electroconvection flows surrounding the dynamic clusters oscillates periodically with $E$. Since dynamic clusters are stabilized by the competition between the external confinement of electroconvective flows and the repulsion between charged bacteria, the oscillation of the magnitude of electroconvective flows results in the instability of the dynamic clusters: The radius of dynamic clusters oscillates in the same frequency of $E$ (Movie 5). Correspondingly, the dynamic clusters transform between an expanding state and a shrinking state periodically (Figs. 2a-b).

In this study, the radius $R$ of the dynamic clusters in the expanding state is measured to characterize the size of the dynamic clusters. As a result of the competing between the electroconvection flows and the electrostatic repulsion, there is a critical size of dynamic clusters. Above the critical size, the dynamic clusters split into smaller ones (Movie 4). To find out the critical $R$ for splitting, the growth of one cluster is followed from the beginning (Movie 9). The radius $R$ where the splitting occurs is recorded as the critical size $R_c$. Alternatively, the distribution of $R$ in the final steady states is measured. Figure 2c represents a typical size distribution obtained at $E = 0.035$ V/$\mu$m and $f = 0.6$ Hz. The distribution is characterized by a peak value $R_p$. However, the distribution is not symmetry: there is a tail on the right of $R_p$. It is found that the maximum size at the end of the tail is highly consistent with the $R_c$ identified by following the growth process. Based on this result, the maximum sizes in the steady distributions are taken as the critical size $R_c$ (Materials and Methods). $R_c$ is sensitive to the frequency: Increasing the frequency gives rise to the decrease of $R_c$ (Fig. 2d). As a reasonable understanding, we suggest that a response time is necessary for the dielectric liquid to establish steady electroconvective rolls. As the response time is longer than $0.5/f$, the electric field will be inverted before the steady convection rolls form. In this case, the maximum magnitude of the convection flows becomes frequency-dependent, giving rise to a frequency-dependent $R_c$. In contrast, $R_c$ does not change significantly as $E$ changes (Fig. 2e). A possible mechanism is that as $E$ enhances the electroconvection, it also enhances the dipolar repulsion between charged particles.

If the critical size results from the balance between electroconvective flows and the electrostatic repulsion inside cluster, it is expected that $R_c$ does not depend on the global concentration of the suspension. However, our observations reveal that as the concentration is higher than $\rho_c = 4.8 \times 10^8$ cfu/ml, the critical size increases sharply with the global concentration (Fig. 2f). Below the concentration $\rho_c$, the number density of dynamic clusters is low and heterogeneous in the system (left inset in Fig. 2f). However, above the concentration of $\rho_c$, the number density becomes much higher (right inset in Fig. 2f) such that the electrostatic repulsions between clusters become significant and provide an additional mechanism to balance the internal repulsion between bacteria. Therefore, at high concentrations, there are two mechanisms, the electroconvective flows and the repulsion between clusters, which balance the electrostatic repulsion between bacteria cells inside the clusters, promising larger critical size $R_c$.

Accompanying the oscillating of $R$, the concentration of the bacteria inside clusters oscillates between the expanding state and the shrinking state (Movie 5). In the shrinking state, the concentration in the clusters is not uniform: It is characterized by a dense shell (Fig. 2b bright region) and a sparse core (Fig. 2b dark region). The understanding is that the shrink of the clusters is induced by the enhancement of the electroconvection flows which produce an external pressure on the surface of the clusters. As a result, the condensation starts from the surface and develop toward the core gradually. Accompanying the condensation, the electrostatic repulsions between bacteria increase. However, the compression lasts only for a finite time (T/4, T is the period of AC) such that the condensation cannot develop into the core of clusters, giving rise to a dense shell and a sparse core. The core-shell structure produces an interface between a high stress domain (dense shell) and a low stress domain (sparse core). A small density fluctuation on the interface will be amplified by the stress difference and leads to the breaking of the shell, giving rise to the splitting of clusters (Figs. 3a-d and Movie 4).

The dependence of $R_c$ on $f$ suggests that the splitting (or merging) of clusters can be triggered (or enhanced) by tuning $f$. To verify this scenario, starting from the steady state at $f = 0.3$ Hz, the frequency is tuned directly to a larger value $f = 0.6$ Hz. As expected, the largest clusters split into smaller ones (Movie 10). By contrast, the decrease of frequency does not lead to the merging of existing clusters. Observations show that as clusters reach the steady state, they don't perform long-distance diffusion. Therefore, the merging of clusters becomes difficult.

Given the concentration of bacteria, the formation of oscillating clusters occurs only in a finite region in the plane of $E$-$f$. Figure 4a presents the phase diagram measured at the concentration of $2.4 \times 10^8$ cfu/ml. Outside the oscillating cluster region, a stronger electric field leads to the directed motion of bacteria such that the charged bacteria are absorbed on the electrodes and becomes damaged. At high frequencies, it is difficult for the dielectric liquid to develop steady and strong convection flows. As a result, a condensed phase can form but further condensation for clusters cannot occur. The interface between the condensed phase and the surrounding sparse phase is steady and no oscillating is observed (Movie 1). This is distinct from the dynamic clusters. Phase boundaries in $E$-$f$ plane are dependent on the global concentration of bacteria. Increasing the concentration shifts the phase boundaries toward low frequencies (Fig. S1a). Increasing concentration also leads to the increase of the peak size and the critical size of dynamic clusters (Fig. S1b). As the concentration is high enough, clusters become jammed (Fig. S1c).

Electroconvective flows form closed rolls across the systems. Therefore, the scale and the magnitude of fluid flows depend on the thickness of the chamber. It is expected that increasing the height of the chamber, which promises larger scale of the convection flows, can improve the critical size of the dynamic clusters at the given $E$ and the concentration of bacteria. To verify this scenario, experiments are conducted with two samples with different heights (the distance between the electrodes) at the same $E$ and bacterial concentration. Figure 4b exhibit the size distributions observed. As expected, the size distribution in the system with doubled thickness shifts significantly toward the large size side. Both the peak size and the critical size increase by two times. These results offer additional evidences that the electroconvective flows play a critical role in forming the dynamic clusters.

**Conclusions**

In summary, bacteria in suspensions subjected to an AC field exhibit interesting modulated collective behavior. They form dynamic clusters oscillating at the same frequency with the electric field. Above a critical size, dynamic clusters split into smaller ones spontaneously. Oscillating and splitting can be well understood based on the mechanism of electroconvective flows. In this mechanism, the biological activity of bacteria is not essential. Instead, the physical properties as charged particles play the central role in forming dynamic clusters. These observations reveal that the collective behaviors of active units in liquid can be tuned and modified by applying control on hydrodynamics. In addition, the formation of dense dynamic clusters significantly improves the local concentration of bacteria. This has potential clinic applications to detect the bacteria early and quickly. Finally, to have a full and complete understanding on the mechanism of electroconvective flows, further observations are critical. In future studies, this has to be explored with more attention.

**Author Contributions**
THZ designed the experimental research; MMB and IZI performed experiments and contribute equally; MMB, KC and THZ analyzed experimental data and wrote the paper.

**Declaration of Interests**

The authors declare no conflict of interest.

**Acknowledgements**

We thank He-Peng Zhang for the help in bacteria cultivation and valuable discussions. T.H.Z. and K.C. acknowledge financial support of National Natural Science Foundation of China (Grant No. 11974255 and 11635002 to T.H.Z.).

**Figures**

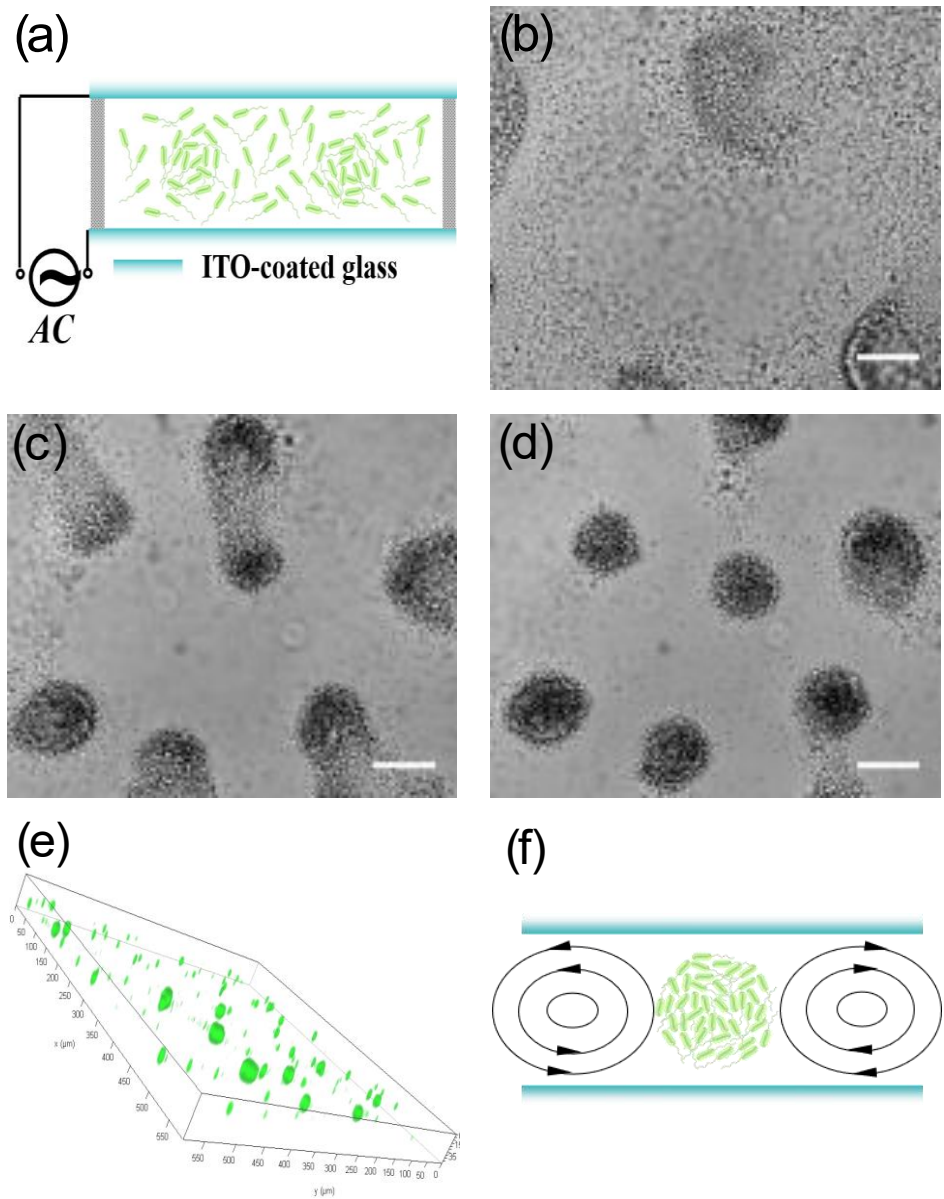

**Fig. 1 Emergence of dynamic clusters.** $E = 0.035$ V/$\mu$m, $f = 0.2$ Hz and $\phi = 2.4 \times 10^8$ cfu/ml. (a) Schematics of the experimental setup. (b) Swimming bacteria are gathering to form dense domains as the electric field is turned on. (c) Emergence of dense dynamic clusters (d) Steady clusters in the final state. (e) Three-dimensional structure of the dynamic clusters observed with confocal microscopy. (f) Electroconvective flows surrounding the dynamic clusters. The direction of rolls changes periodically with the AC field. Scale bar: 20 $\mu$m.

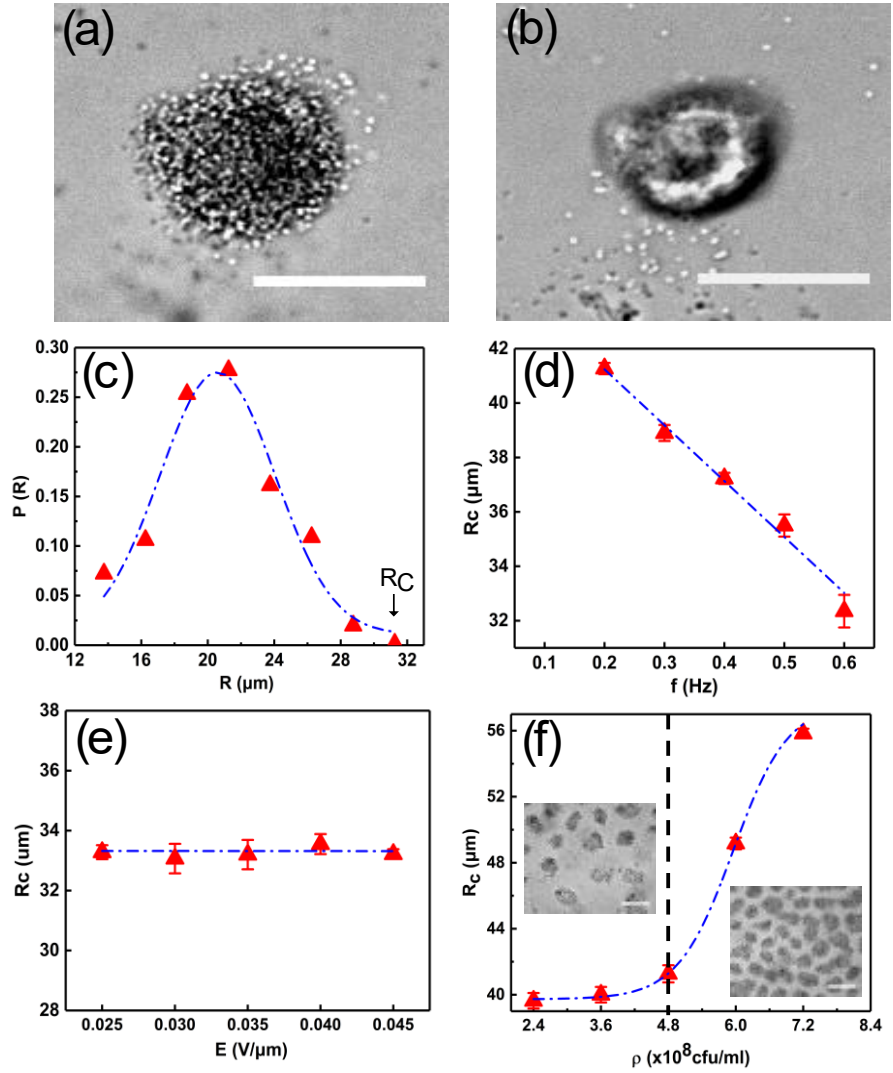

**Fig. 2 Critical size of dynamic clusters.** (a-b) Expanding and shrinking states at $E$ = 0.035 V/$\mu$m and $f$ = 0.2 Hz. Scale bar: 20 $\mu$m. (c) Size distribution in the final steady state at $E$ = 0.035 V/$\mu$m and $f$ = 0.6 Hz. (d) Dependence of critical size on frequency at $E$ = 0.035 V/$\mu$m. (e) Effect of $E$ on critical size at $f$ = 0.45 Hz. (f) Critical size of clusters formed at different concentrations with $E$ = 0.03 V/$\mu$m and $f$ = 0.45 Hz. Scale bar: 50 $\mu$m.

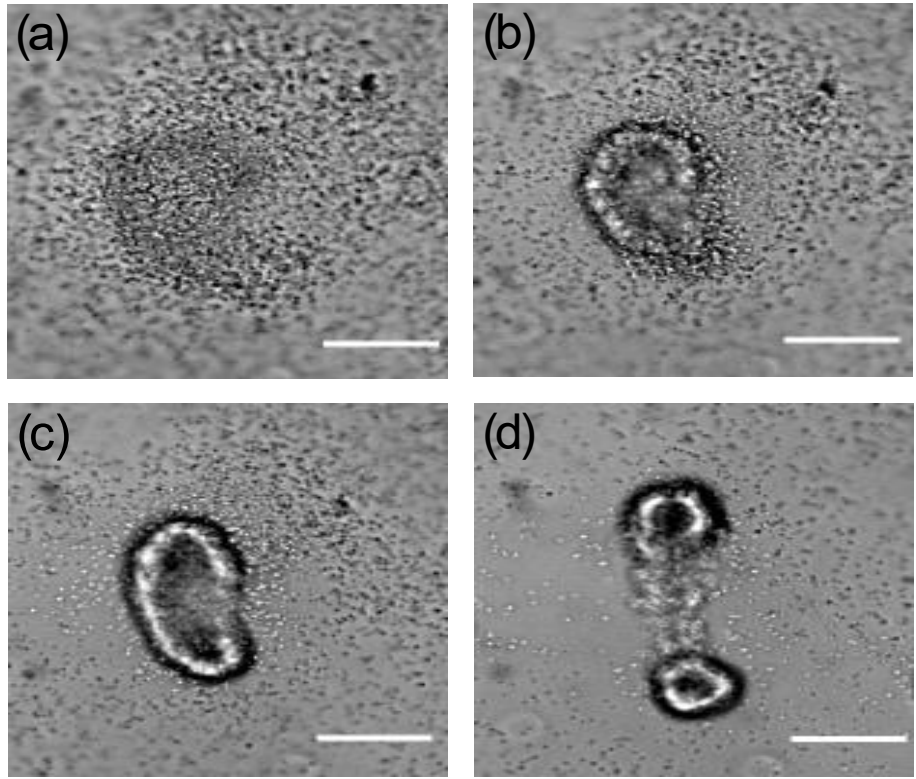

**Fig. 3 Formation and splitting.** $E = 0.035$ V/$\mu$m and $f = 0.2$ Hz. (a) Early stage of the dynamic clusters. (b) Dense shell forms as the concentration in the growing clusters is high. (c) Distorted core-shell structure. (d) Two smaller clusters forms after the splitting. Scale bar: 20 $\mu$m.

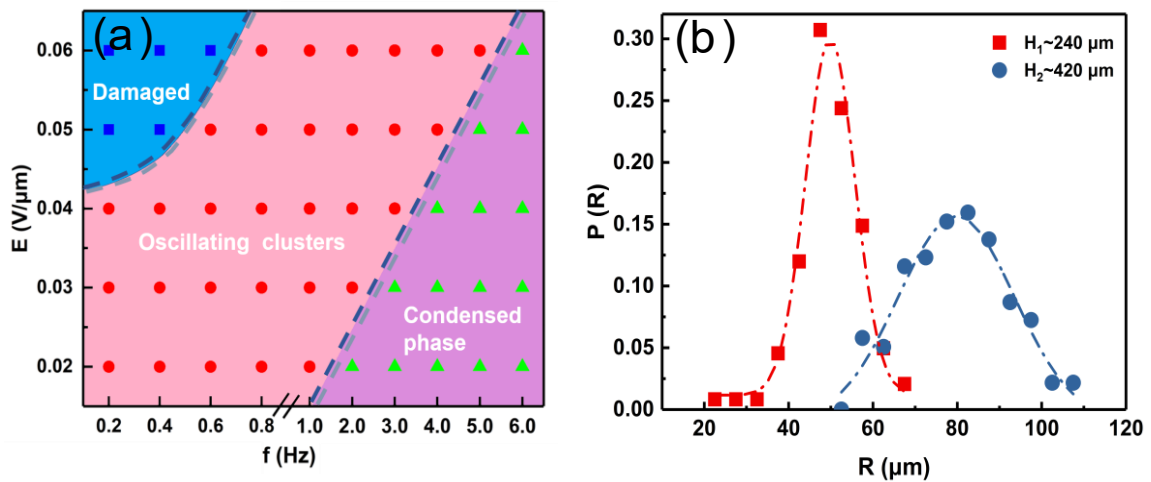

**Fig. 4 Phase diagram and the effect of height on size distribution.** (a) Different structures that form under an AC field at bacterial concentrations of 2.4x10$^8$ cfu/ml. The dashed lines are lines where qualitative changes either in structure or dynamics are observed. (b) Size distribution of clusters formed at different height. $E = 0.035$ V/$\mu$m, $f = 0.3$ Hz, $H_1 = 240$ $\mu$m, $H_2 = 420$ $\mu$m.